\documentstyle[prl,aps,epsf,multicol]{revtex}
\begin{document}

%
\newcommand{\fig}[2]{\epsfxsize=#1\epsfbox{#2}}
%
%
\newcommand{\passage}{
\end{multicols}\widetext\noindent\rule{8.8cm}{.1mm}%
  \rule{.1mm}{.4cm}} 
 \newcommand{\retour}{
\noindent\rule{9.1cm}{0mm}\rule{.1mm}{.4cm}\rule[.4cm]{8.8cm}{.1mm}%
         \begin{multicols}{2} }
 \newcommand{\unecol}{\end{multicols}}
 \newcommand{\deuxcol}{\begin{multicols}{2}}
%
%
\newcommand{\beq}{\begin{equation}}
\newcommand{\eeq}{\end{equation}}
\newcommand{\beqa}{\begin{eqnarray}}
\newcommand{\eeqa}{\end{eqnarray}}
\newcommand{\eqbreak}{
\end{multicols}
\widetext
\noindent
\rule{.48\linewidth}{.1mm}\rule{.1mm}{.1cm}
}
\newcommand{\eqresume}{
\noindent
\rule{.52\linewidth}{.0mm}\rule[-.1cm]{.1mm}{.1cm}\rule{.48\linewidth}{.1mm}
\begin{multicols}{2}
\narrowtext
}
%

\author{Horacio Castillo and Pierre Le Doussal} 

\address{CNRS-Laboratoire de Physique Th{\'e}orique de 
l'Ecole Normale Sup{\'e}rieure,
24 rue Lhomond,75231 Cedex 05, Paris France.}

\title{Freezing of dynamical exponents in low dimensional random media} 
\date{\today}
\maketitle

\begin{abstract}
A particle in a random potential
with logarithmic correlations in dimensions $d=1,2$
is shown to undergo a dynamical transition at $T_{dyn}>0$.
In $d=1$ exact results demonstrate that $T_{dyn}=T_c$, the static
glass transition temperature, and that the dynamical exponent 
changes from $z(T)=2 + 2 (T_c/T)^2$ at high temperature 
to $z(T)= 4 T_c/T$ in the glass phase. The same formulae
are argued to hold in $d=2$. Dynamical freezing 
is also predicted in the 2D random gauge XY model and related systems.
In $d=1$ a mapping between dynamics and statics is unveiled and
freezing involves barriers as well as valleys.
Anomalous scaling occurs in the creep dynamics.
\end{abstract}


\deuxcol

The motion of topological defects by thermal activation over pinning barriers determines
the slow glassy dynamics in numerous disordered systems, e.g.
domain walls in dirty magnets, vortices in superconductors,
dislocations in pinned lattices
\cite{blatter}. While treating many interacting
extended defects in the
presence of disorder remains a major challenge,
progress can be made on the simpler, already 
non trivial problem of a single point defect. 
A model of great interest is
a particle diffusing in a random potential with
log-correlations, i.e 
barriers growing logarithmically with scale. In 2D 
it precisely describes a single vortex in a XY spin model
with gaussian random gauge disorder \cite{nattermann},
and is relevant to a host of related systems
e.g. vacancies in pancake lattices of layered 3D superconductors \cite{baruch,blatter},
dislocations in 2D lattices with smooth disorder \cite{dc_pld1},
electrons on helium \cite{tito}. As a prototype model of diffusion in 
a complex phase space or in random media, it is of broader interest to
relaxation in glasses \cite{goldstein}, transport in solids
\cite{alexander} population biology \cite{dahmen},
non hermitian quantum mechanics \cite{chalker} and vortex glass
dynamical scaling \cite{ledou_vino}.
Similar models where used to study the dynamical generation 
of broad (e.g. power law) distributions of
relaxation times and its relation to aging \cite{laloux}.
Its static limit also appears in Integer Quantum Hall (QH) transition
studies~\cite{REF:static_space} and its quantum extension 
in QH bilayer systems~\cite{fisher}.

Early studies of this model, in the context of tracer diffusion in 2d potential flows,
used RG methods perturbative
in the disorder strength $\sigma$
\cite{REF:kly,REF:bcgl,honkonen}. The position $x(t)$
of a particle (in any $d$) satisfies a Langevin equation:
\begin{eqnarray}
\dot{x}(t)= - \nabla V(x(t)) + \eta(t)
\end{eqnarray}
$\langle \eta(t) \eta(t') \rangle = 2 T \delta(t-t')$ 
being a thermal noise ($\overline{X}$ and 
$\langle X \rangle$ respectively denote
disorder and thermal averages).
It was found that if 
correlations grow logarithmically:
\begin{eqnarray}
\Delta(x-x') = \overline{(V(x) - V(x'))^2} 
\sim 4 \sigma \ln |x-x'|
\label{EQ:log_corr} 
\end{eqnarray}
the diffusion is anomalous with a temperature $T$ dependent
``dynamical exponent'' $z$
given by \cite{REF:bcgl,honkonen}:
\begin{eqnarray}
\overline{\langle x(t)^2 \rangle} \sim t^{2/z} \quad , \quad
z = 2 + 2 \frac{\sigma}{d T^2}
\label{EQ:zht}
\end{eqnarray}
Strikingly, this result for $z$ was conjectured \cite{REF:bcgl} to hold to 
{\it all orders} in $g = \sigma/d T^2$. This 
was confirmed in $d=1$ and $d=2$ by general arguments
and a three loop calculation \cite{honkonen_pismak} 
(which also found $O(g^3)$ corrections to $z$ in $d \geq 3$).
Further support came from excellent agreement with simulations 
\cite{deem}, performed in $d=2$ for $0< g < 0.8$, and exact results
in $d=1$ \cite{ledou_vino}
for the velocity $v \sim f^{z-1}$ at small applied force $f$,
with $z$ again as in (\ref{EQ:zht}).

Recently, however, the {\it statics} of this model has been 
investigated in several works in $d=2$
\cite{nattermann,REF:static_space,REF:static_tree,REF:kor_nat,scheidl}
and any $d$ \cite{carpentier_ledou}.
As a result we now know that there is in fact 
{\it a transition} at $T=T_c=\sqrt{\sigma/d}$ to a strong disorder, low $T$ glass phase.
This glass phase appears non trivial, as dominated by {\it a few} states,
reminiscent of replica symmetry breaking (RSB).
It is related, approximately, to the Random Energy Model (REM) \cite{rem}
and, more closely, to
the directed polymer on a 
Cayley tree (DPCT) \cite{dpct}. It is thus
an outstanding problem to investigate whether 
this equilibrium transition has a dynamical counterpart, and how it can be
compatible with (\ref{EQ:zht}).

In this Letter we solve the apparent paradox. In $d=1$ we demonstrate
that there is indeed a dynamical transition at the static $T_c$ and
obtain the dynamical exponent $z(T)$ at all $T$. This is achieved through
exact results and a real space RG (RSRG) method. Simple
arguments and bounds also show that a dynamical transition occurs in $d=2$,
with analogous behaviour of $z(T)$. A similar dynamical freezing is also
predicted in the 2d random gauge XY model, since
the XY phase and its boundary are dominated 
by the single (i.e dilute) vortex limit
\cite{nattermann,carpentier_ledou}.

We first give a simple argument indicating that
the $1/T^2$ divergence in (\ref{EQ:zht})
{\it cannot} hold at low temperature. First, 
{\it in any given sample} (in finite $d$), the  
characteristic time $t$ for a particle to
escape a region of size $L$ satisfies:
\begin{eqnarray}
\ln t \approx B/T \leq [V_{max}(L)-V_{min}(L)]/T \equiv B_{max}/T
\label{bound0}
\end{eqnarray}
i.e is given as $T \to 0$ by the Arrhenius estimate
where $B$ is the energy barrier encountered by the particle, obviously 
bounded by 
the difference $B_{max}$ between the absolute maximum $V_{max}(L)$ and 
minimum $V_{min}(L)$ of the potential in the region.
Second, we now know~\cite{REF:kor_nat,REF:static_space,scheidl,carpentier_ledou}
that these extremal values satisfy
$V_{min}(L) = - 2 \sqrt{\sigma/d} (d \ln{L} - \frac{1}{2} 
\gamma \ln \ln L) + \delta V$
where $\gamma$ is a universal number and 
$\delta V$ has $O(1)$ sample to sample fluctuations.
Thus, defining $z \equiv \ln t /\ln L$ at large $L$ \cite{FNOTE:L_x} this 
yields, in any dimension $d$, the bound:
\begin{eqnarray}
z \leq 4 \sqrt{\sigma d}/T  \qquad T \to 0.
\label{bound}
\end{eqnarray}
Since (\ref{EQ:zht}) is exact perturbatively 
to all orders in $g=\sigma/T^{2}$ in $d=1,2$, (\ref{bound}) 
{\it implies} that a dynamical transition
{\it must} occur in $d=1,2$ at a finite 
temperature $T_{dyn}(d)>0$.

In $d=1$, which we now study, it is natural
to guess that the upper bound in
(\ref{bound}) gives the exact $z(T)$ for all $T<T_c$.
Indeed one expects that the Arrhenius law (\ref{bound0}) should hold
in the energy dominated glass phase and
furthermore $B$ should equal its upper bound in (\ref{bound0})
since in $d=1$ there is only a single path.
We now confirm this with analytical results at all $T$
using the first passage time
approach~\cite{REF:gor_blat}. We also show that (i) in $d=1$ there is
a direct correspondence between dynamical (e.g.
exponents) and static quantities (ii)
the dynamical exponent $z$ can be unambiguously defined.
Our conclusions being independent of boundary conditions
we choose them reflecting at $x=0$ \cite{REF:gor_blat}
and define $t$ as the first passage time at site
$x=L$ of a particle which starts at $x=0$.
Some of our results are valid for {\em any} potential landscape
but are mostly applied to gaussian log-correlated
potentials (\ref{EQ:log_corr}).

As shown below it is sufficient to focus,
in any given disorder realization $V(x)$,
on the (thermal) {\it mean first passage time}
$t_{1}(\{V\}) \equiv \langle t \rangle \equiv \tau$. We
first obtain its {\it typical} behaviour (i.e in any sample),
its large deviations being studied later. The exact formula for 
$t_{1}(\{V\}) $ is~\cite{REF:gor_blat}:
\begin{equation}
t_{1}(\{V\}) = \frac{1}{T} \int_{0}^{L} \!\! dy \! \int_{0}^{L} \!\! dx 
\; \theta(y - x) \; {\rm e}^{[V(y) - V(x)]/T} ,
\label{EQ:tp_1}
\end{equation}
where $\theta(x)$ is the step function. Since the 
canonical partition function for the same disorder realization is
$Z_{L}(\{V\}) \equiv \int_{0}^{L} \!\! dx \; \exp(- V(x)/T)$,
(\ref{EQ:tp_1}) strongly resembles {\it two copies of the statics},
one with disorder $+V(x)$ (dominated by minima at low $T$),
the other with $-V(x)$ (dominated by maxima of $V$ at low $T$).
Dynamics and statics are thus directly connected by: 
$\ln\{ T [ t_{1}(\{V\}) \!+\! t_{1}(\{-V\}) ]\} = \ln{
Z_L(\{V\})} \!+\! \ln{ Z_L(\{-V\})}$ and also
\begin{eqnarray}
\ln Z_{L/2}(\{-V\}) & \!+\! & \ln Z_{L/2}(\{V\}) \leq 
\ln\{T t_{1}(\{V\}) \}  \nonumber \\
& \! \leq \! & \ln Z_{L}(\{-V\}) + \ln Z_{L}(\{V\}). \label{EQ:Z_t_1}
\end{eqnarray} 
If the ratio of upper to lower bound converges to unity in
the thermodynamic limit, they uniquely determine the long time
behavior in terms of the statics. This is the case for
gaussian potentials with correlations growing at the most as a power of
the logarithm of the scale. If the correlations do not grow faster
than logarithmically with scale, then $\ln Z_{L}(\{V\}) \sim \ln L$.
If, additionally, the intensive free energy $f=F_L(\{V\})/\ln L 
= - T \ln{ Z_{L}(\{V\})}/\ln L$
is self-averaging, the dynamical exponent $z$ is obtained as:
\begin{eqnarray}
z(\{V\}) \equiv \lim_{L \to \infty} \ln{ t_{1}(\{V\}) }/\ln{L} = -2 f/T, 
\label{EQ:z1_V_def}
\end{eqnarray}
and is also self averaging. This is the case both for uncorrelated
and log-correlated gaussian potentials. For the
former (\ref{EQ:z1_V_def}) gives the normal diffusion
value $z=2$. For the latter, using the known results for $f$ 
\cite{REF:static_space,carpentier_ledou} we obtain our main result
for the dynamical exponent:
\begin{eqnarray}
&& z(T) = z_A(T) \equiv 2 \left( 1+\sigma/T^2 \right) 
\qquad ( \mbox{for} \; T \geq T_{\rm c} \def \sqrt{\sigma} )
\nonumber \\
&& z(T) =  4 \sqrt{\sigma}/T 
~~~ ~~~~~~ ~~~~~ ~ \qquad ( \mbox{for} \; T \leq T_{\rm c} ).
\label{EQ:z_T_V}
\end{eqnarray}
Thus a dynamical transition, away from the ``annealed'' value $z_A(T)$
given by (\ref{EQ:zht}), occurs at the same temperature
$T_{dyn}=T_{\rm c} = \sqrt{\sigma}$ as the equilibrium transition.
At $T_c$ freezing occurs in the 
thermal configurations which dominate $\langle t \rangle$, simultaneously
around minima and maxima of the potential (i.e in
the two copies), thus in the effective barrier.
Interestingly, this transition coincides with the onset of
logarithmic corrections. Indeed one can also 
characterize {\it typical} finite size fluctuations.
Using that $F_L=f (\ln L - \frac{1}{2} \gamma(T) \ln (\ln L)) + \delta F$
where $\delta F$ has $O(1)$ sample to sample
fluctuations \cite{carpentier_ledou}
we find that $t_{1}(\{V\})/\tau_{typ}$ 
has a well defined $O(1)$ limit distribution
at large $L$, the {\it typical} 
mean first passage time
being $\tau_{typ} = L^{z(T)} (\ln L)^{- \alpha(T)}$
with $\alpha(T) = 2 \gamma(T) \sqrt{\sigma}/T$,
$\gamma(T)=0$ for $T>T_c$ but $\gamma(T_c)=\frac{1}{2}$
and $\gamma(T)=\frac{3}{2}$ for $T<T_c$.
For faster
growing (e.g. power law) correlations,
the ratio between the bounds in
Eq.~(\ref{EQ:Z_t_1}) does not converge to one, and 
even the leading order of $\ln t_{1}(\{V\})$  
still fluctuates at large $L$ \cite{REF:gor_blat}
as in the unbiased Sinai model \cite{REF:RSRG_Sinai}.

We now show as promised that $z$
can be defined unambiguously from the {\it mean} first passage time alone.
In principle one could define a full set of dynamical exponents 
$z_{p}(\{V\}) = \lim_{L \to \infty} \ln{ t_{p}(\{V\}) }/p \ln{L}$
from higher thermal moments $t_{p}(\{V\}) \equiv \langle t^{p} \rangle$.
Using the expression~\cite{REF:gor_blat}:
\begin{eqnarray}
t_{p}(\{V\}) = \frac{p!}{T^{p}} \int_{0<x_i,y_i<L} \prod_{i=1}^{p}
\theta_{y_{i},x_{i}} \theta_{y_{i},x_{i-1}}
{\rm e}^{\frac{V(y_{i}) - V(x_{i-1})}{T}} \nonumber
\end{eqnarray}%
and $\theta_{x,y}\equiv \theta(x\!-\!y) \leq 1$ we obtain,
comparing with (\ref{EQ:tp_1}),
$\ln{ t_{p}(\{V\}) } /p \ln{L}  \leq \ln{ t_{1}(\{V\})
}/ \ln{L}  + \ln{ p!} / p \ln{L}$,
which together with the general inequality $ \langle t \rangle^{p}
\leq \langle t^{p} \rangle $ leads to:
\begin{equation}
z_{p}(\{V\}) = z_{1}(\{V\}) \qquad 
\mbox{and} \qquad \overline{z_{p}} = \overline{z_{1}}. 
\label{EQ:zp_z1_V}
\end{equation}
for any integer $p \ge 1$. This is because
the thermal distribution of $t$ has exponential decay
and all moments $p \ge 1$ are controlled at large $L$ by the
{\it largest} relaxation time in a given sample.
Thus $z = z_{1}$ characterizes the dynamics.

By contrast, the distribution of escape times $P_L(\tau)$ 
with disorder realization has broad tails extending in
the region $\tau \gg \tau_{typ} \sim L^{z(T)}$, where we
obtained the estimate:
\begin{eqnarray}
&& P_L(\tau) d\tau \approx \frac{d\tau}{\tau}
(\frac{\tau_{typ}}{\tau})^{\mu} 
\exp(- \mu^2 \frac{\ln^2(\tau/\tau_{typ})}{8 \ln L} ) .
\label{largedev}
\end{eqnarray}
Here $\mu=T/T_c$ and (\ref{largedev})
is valid for (i) $T<T_c$ and (ii) for $T>T_c$ and 
$\tilde{z} \equiv \ln \tau /\ln L \ge 4$. For $T>T_c$
and $z(T) < \tilde{z} < 4$ one has simply 
$P_L(\tau) \approx \tau_{typ}^{-1} (\tau_{typ}/\tau)^{1+\mu^2}$.
This corresponds to a quadratic (plus linear) multifractal spectrum for 
rare occurrences of $\tilde{z}$. (\ref{largedev})
can be obtained by a Kosterlitz RG 
analysis of (\ref{EQ:tp_1}), as in \cite{carpentier_ledou}.
The result is a non linear Kolmogorov equation for 
$P_L(\tau)$ as a function of $\ln L$,
identical to the one describing the partition sum of
two directed polymers on a Cayley tree, seeing opposite disorder,
with constrained endpoints $x<y$. It yields 
(\ref{largedev}) up to log corrections, neglected
here. More empirically (\ref{largedev}) can be obtained
from the moments:
\begin{eqnarray}
\overline{{\langle t \rangle}^{p}}
& = & \frac{1}{T^{p}} \int_{0}^{L} 
\prod_{i=1}^{p} \left[ dy_{i} dx_{i} \theta(y_{i} - x_{i}) \right]
\exp \Big\{ \sum_{i,j=1}^{p}
\big[\Delta(y_{i}-x_{j}) \big. \Big.
\nonumber \\
&& \Big. \big. - \frac{1}{2}\Delta(y_{i}-y_{j}) -
\frac{1}{2}\Delta(x_{i}-x_{j}) \big] /2 T^{2} \Big\} .
\label{EQ:t_p_avg}
\end{eqnarray}
which reads as a partition sum for  $2p$ particles,
$p$ of type $y$ (representing hills in the potential landscape) and
$p$ of type $x$ (representing valleys). Same type particles
attract via a potential $\Delta(r)/2T$ while those of opposite type
repel via $-\Delta(r)/T$. This is similar to estimating 
$\overline{Z^{p}}$ in the statics, except that
there only one kind of particles (representing
valleys) appears. Here hills and valleys play symmetric roles.
(\ref{EQ:t_p_avg}) can be estimated for log-correlated potentials and 
for integer $p \ge 1$
as follows. At high $T$ an "entropic" 
variational saddle point dominates (with all
$y$'s and $x$'s far away $O(L)$ from each other). At low $T$ 
an "energetic" saddle point dominates (all $y$'s close
together within $O(1)$, all $x$'s close together, $y$'s and $x$'s far away).
This yields the large $L$ behaviour
(universal since unaffected by changes in $\Delta(r)$ for small
$r$):
\begin{eqnarray}
\overline{{\langle t \rangle}^{p}} 
\sim & L^{2p(1+\sigma/T^2)} 
& \qquad \mbox{for} \quad T \geq T_{\rm c, p}, \nonumber \\ 
\sim & L^{2p(1/p + p\sigma/T^2)} 
& \qquad \mbox{for} \quad T <    T_{\rm c, p}.
\label{EQ:tp_transitions}
\end{eqnarray}
i.e, as for $\overline{Z^{p}}$ in the statics
(of this model
~\cite{REF:static_tree,REF:static_space}, the REM and
the DPCT) there is a sequence of transition temperatures 
$T_{\rm c, p} = \sqrt{p} \, T_{\rm c}$ for the moments.
One can check via a saddle point calculation
of $\overline{{\langle t \rangle}^{p}}=\int d\tau \tau^p P_L(\tau)$
that (\ref{EQ:tp_transitions}) is consistent with (\ref{largedev})
and that the $T_{\rm c, p}$ correspond to a change of behaviour
from rare events to typical events dominance
as the saddle point crosses $\tilde{z}_{sp}=4$.
By analytically continuing $\overline{{\langle t \rangle}^{p}}$ 
as $p \to 0$ one can recover (\ref{EQ:z_T_V}) and the transition
at $T_c$ in typical behaviour. The "entropic" saddle point 
still dominates for $T>T_c$, but the "energetic" saddle point is
replaced at $T<T_c$ by a one-step RSB ansatz. Each kind of
"{particle}" is arranged in $p/m$ groups of $m$ particles
close by in space, while different groups are far away,
with $m=T/T_c$ at the optimum, extending the
static \cite{scheidl,carpentier_ledou} and
the REM and DPCT replica calculations 
\cite{rem}. Compared to 
conventional static RSB which only involves
``valleys'' the interesting feature here is a {\it nonequilibrium RSB } which 
also involves ``hills'': indeed,
near degeneracies of distant barriers result, in the glass phase,
in a {\it nonequilibrium} splitting of the thermal distribution
of the diffusing particle into {\it a few} packets in a single environment
\cite{footnoteoverlap}. These features are
absent for weaker correlations (high $T$
"entropic" saddle) or stronger ones \cite{REF:gor_blat}, where an
"energetic" saddle without RSB dominates (degeneracies are 
subdominant in the Sinai landscape 
except in the presence of a bias \cite{REF:RSRG_Sinai}).

If an external force is applied, the creep velocity $v$ relates to the 
{\it disorder averaged} mean escape time of 
regions of size $L_0 \sim 1/f$ and is thus controlled
by the annealed exponent $z_A(T)$, distinct from $z(T)$
at low $T$, a striking breakdown of naive dynamical scaling.
Freezing manifests itself in large 
finite size corrections to $v$ due to undersampling
of the disorder average. From \cite{ledou_vino},
$v^{-1}= \frac{1}{L} \langle t \rangle $ holds for $f L \gg 1$,
where $\langle t \rangle$ is given by (\ref{EQ:tp_1})
in the tilted landscape $V(x) - f x$. It can be 
approximated as
the average over $N$ samples of sizes $L_0$ of the
escape time in each sample, each distributed with $P_{L_0}(\tau)$.
A saddle point estimate shows that below 
$T_c$, $v \sim f^{z_A-1}$ for $y = - \ln L/\ln f > y^*= 2/\mu^2-1$,
but that for smaller sizes $1<y<y^*$, the typical
$v \sim f^{z_m-1}$ where the exponent $z_m= \frac{4 T_c}{T}
\sqrt{(1+y)/2} - y + 1$ smoothly interpolates between the annealed and
quenched one.

The RSRG method, previously devised
to describe diffusion in the Sinai landscape
(for details see~\cite{REF:RSRG_Sinai}) and in
a broader class \cite{REF:RSRG_general}, allows to obtain
complementary information, e.g. about 
distribution of positions. Here, in the log-correlated landscape,
it is implemented numerically.
From the original set of
(i) alternating local extrema $V(x_i)$,
(ii) their energy differences (``barriers''
of heights $F_i=|V(x_i)-V(x_{i+1})|$)
and (iii) the segments between them (``bonds'' of lengths $\ell_i=x_{i+1}-x_i$),
one constructs iteratively the ``renormalized landscape at 
scale $\Gamma$'' by removing as $\Gamma$ increases all barriers between 
$\Gamma$ and $\Gamma + d \Gamma$
and merging the corresponding bonds.
This decimation retains only the large barriers and deep valleys.
The Arrhenius dynamics of a particle starting at $x_0$ is 
\begin{figure}[thb] 
\centerline{\fig{7cm}{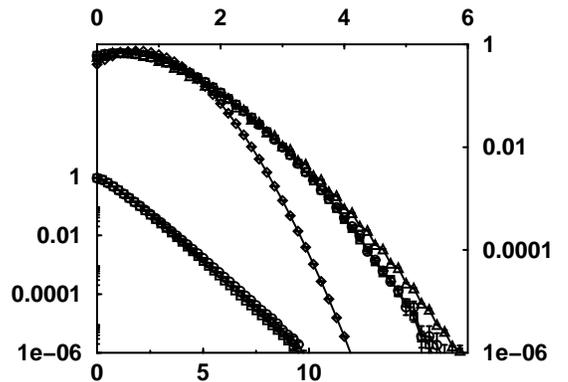}}
\caption{\label{FIG:P_scale} \narrowtext 
Probability distributions for $\sigma=\frac{1}{2}$:
(i) without and (ii) with additional short range disorder.
Upper right:
$Prob_{\Gamma}((F-\Gamma)/F^{\rm typ})$ for $2^{23}$ sites
(i) with 2700 samples, initial ($\Gamma = 0$, triangles) and 
asymptotic ($\Gamma = 12$, squares) (ii) with 1000 samples
initial ($\Gamma = 0$, diamonds) and asymptotic
($\Gamma = 18$, circles). Lower left:
asymptotic $q(X=x/\overline{\ell}_\Gamma)$; (i) 
($\Gamma = 12$, squares)
(ii) ($\Gamma = 18$, circles). }
\end{figure}

\begin{figure}[thb] 
\centerline{\fig{6cm}{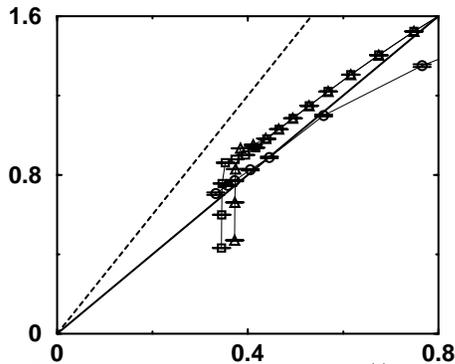}}
\caption{\label{FIG:alpha_gamma} \narrowtext
On same graph: dependence of (i) maximum barrier
$B_{max}=V_{max}-V_{min}$ on system size $L$ (ii) 
minimum barrier $B=\Gamma$ left at decimation scale $\Gamma$
on average bond length $\overline{\ell}_\Gamma$. Vertical axis: 
$4 - B/(\sqrt{\sigma} \ln R)$, horizontal axis: $ \ln (4 \pi \ln R) / \ln R$
with (i) $B=B_{max}$, $R=L$ (circles) and (ii) $B = \Gamma$, 
$R=\overline{\ell}_\Gamma$, $L = 2^{23}$ (squares) and
$L = 2^{21}$ (triangles).
A graph going to the origin means $B \sim 4 
\sqrt{\sigma} \ln R$ for $R \to \infty$. Finite size 
corrections $B=4 \sqrt{\sigma} (\ln R 
- \frac{1}{2} \gamma \ln (4 \pi \ln R))$ with $\gamma=3/2$
(dashed line) 
$\gamma=1$ (full line) are shown.}
\end{figure}%
approximated by putting it at time $t$ 
at the bottom of the bond which contains $x_0$ 
in the renormalized landscape at $\Gamma = T \ln t$.
The errors are small if the distribution of
renormalized barriers is broad compared to $T$ (infinitely
broad in Sinai). Since here barriers remain finite, the method
should be exact only as $T/T_c \to 0$, i.e when the
thermal packet is concentrated in a single well. Perturbative
corrections in $T/T_c$ can be computed
by considering the small occupation probability of neighboring secondary
wells. In practice, the method works surprisingly better
and gives the exact $z(T)$ up to $T_c$.

We have found that under renormalization 
the probability distributions for rescaled variables reach
fixed point forms, namely 
$\text{Prob}_{\Gamma}((F-\Gamma)/F_\Gamma^{\rm typ}) 
\to  P^*((F-\Gamma)/F^{\rm typ})$
for rescaled barriers (Fig.~\ref{FIG:P_scale}) and 
$\text{Prob}_{\Gamma}(\ell/\overline{\ell}_\Gamma) 
\to  Q^*(\ell/\overline{\ell}_\Gamma)$.
Here $F_\Gamma^{\rm typ}$ flows to a constant
$F^{\rm typ} \approx 4.4 \sqrt{\sigma}$ and $\overline{\ell}_\Gamma = L/N_\Gamma$ 
is the average bond length. Since two barriers are 
decimated at each step the number of remaining barriers $N_\Gamma$
satisfies $\partial_\Gamma N_\Gamma = -2 \alpha_\Gamma N_\Gamma$
where $\alpha_\Gamma \equiv Prob_\Gamma(F=\Gamma) \to \alpha^*
=P^*(0)/F^{\rm typ}$, a constant. Thus
the bond length grows as $\overline{\ell}_\Gamma
\sim \exp(2 \alpha^* \Gamma) \sim t^{1/z}$ and using
$\Gamma = T \ln t$ one recovers the
dynamical exponent $z = 1/(2 \alpha^* T)$. Numerically
we find $1/(2 \alpha_\Gamma) \approx 4 \sqrt{\sigma} 
(1 - 0.25 \ln( 4 \pi \ln \overline{\ell}_\Gamma) / 
\ln{\overline{\ell}_\Gamma})$
and thus a value of $z$ consistent with (\ref{EQ:z_T_V}).
The diffusion front, computed as in \cite{REF:RSRG_Sinai}, converges
to a scaling form as
$\overline{Prob_\Gamma(x t|00)} \to \overline{\ell}_\Gamma^{-1}
q(X=x/\overline{\ell}_\Gamma)$, represented in 
Fig.~\ref{FIG:P_scale}. $\overline{\ell}_\Gamma$ 
is thus the only relevant lengthscale, all moments
of the displacement scaling as 
$\overline{\langle x(t)^{k} \rangle} \sim \overline{\ell}^{k}_\Gamma \sim
t^{k/z}$. Finally we obtained good numerical
evidence (see Fig.~\ref{FIG:P_scale})
that all above asymptotic scaling functions, as well as $\alpha_0\equiv \sqrt{\sigma}
\alpha^*$ and $F_0 \equiv F^{\rm typ}/\sqrt{\sigma}$, do not change
upon adding short-range disorder and are thus {\it universal}
($\overline{\ell}_\Gamma$ does change by a constant factor.)

We now address $d \ge 2$. First we note that the bound (\ref{bound}) 
can be improved to $z(T) \leq 2 \sqrt{d \sigma}/T$ as
$T \to 0$ for any $d \ge 2$. Indeed, in order to escape 
the particle now only needs to find a set of 
saddles which connects to the boundary. A percolation 
and counting argument shows that it can do so by
remaining within sites such that $V(x)/\ln L \leq 0$.
Thus the relevant barrier is bounded as $B \leq - V_{min}$.
In $d=2$ this bound is likely to be saturated
since the particle still finds deepest minima,
yielding $z = 2 \sqrt{2 \sigma}/T = 4 T_c/T$.
Since this expression matches (\ref{EQ:zht})
at the static $T_c$ in $d=2$, a likely scenario 
is that $T_{dyn}=T_c$ and that the expression holds for all $T<T_c$
\cite{other}.
A single vortex in a random gauge XY model will 
experience a similar dynamical freezing.

To conclude we demonstrated dynamical transitions in $d=1,2$.
In $d=1$ we found anomalous scaling of the creep velocity,
novel freezing phenomena involving
barriers and a finite $T$ generalization of Arrhenius law
$\tau \!\sim\! e^{-2 F_L/T}$. Extensions will appear elsewhere.

\unecol


\begin{references}
\bibitem{FNOTE:L_x}
equivalent to definition (3) if the maximum excursion $L$
is proportional ${\overline{\langle x(t)^2 \rangle}}^{1/2}$,
as checked below via RSRG.
\bibitem{other} If present at all, deviations are constrained to be small.
\end{references}
\end{document}